\def\BH{black hole}

\def\PRD{Phys. Rev. D}
\def\PRL{Phys. Rev. Lett.}
\def\GB{\text{GB}}

\def\pano{\newline}
\def\sch{Schwarzschild}

\documentstyle[eqsecnum,aps,prd,preprint]{revtex}
\begin{document}
\preprint{
\vbox{
\rightline{CERN-TH/97-116}
\rightline{hep-th/9708014}
}}
\draft

\title{Entropy and topology for gravitational instantons}
\author{Stefano Liberati$^1$ 
\thanks{Electronic address: liberati@sissa.it} 
and Giuseppe Pollifrone$^{2,3}$
\thanks{Electronic address: giuseppe.pollifrone@cern.ch}
}

\address{$^1$ Scuola Internazionale Superiore di Studi Avanzati, 
Via Beirut 2-4, 34013 Trieste, Italy} 
\address{$^2$ Theory Division, CERN, CH-1211 Geneva 23, Switzerland}

\address{$^3$ Dipartimento di Fisica, Universit\`a di Roma ``La
Sapienza,"  
and INFN, Sezione di Roma, \\
Piazzale Aldo Moro 2,
00185 Roma, Italy}
\maketitle
\begin{abstract}
In this work a relation between topology and thermodynamical
features of gravitational instantons 
is shown. The expression for the Euler characteristic, through the 
Gauss--Bonnet integral, and the one for the entropy of gravitational 
instantons are proposed in a form that makes the relation between them 
self-evident. A new formulation of the Bekenstein--Hawking 
formula, where the entropy and the Euler characteristic are related by 
$S=\chi A/8$, is obtained.
This formula provides the correct results for 
a wide class of gravitational instantons described by both
spherically and axially symmetric metrics.
\end{abstract}

\section{Introduction}
\label{sec:intro}
At the beginning of the seventies an impressive series of theoretical 
results in General Relativity were achieved, which led to an interpretation of 
some laws of black hole physics as thermodynamical ones \cite{BCH,C,CR}. 
Remarkably, it was found that \BH s were endowed with an intrinsic entropy 
proportional to the horizon area, the so-called Bekenstein--Hawking
entropy. Consistency for such a
framework was subsequently achieved thanks to Hawking's discovery of black
hole radiation\cite{Haw}, which results from the application of 
quantum field theory to such peculiar space-times. 
Soon after, thermodynamical aspects of black holes appeared
more evident in the Euclidean path-integral approach\cite{GH}.
If one considers the Einstein--Hilbert action plus a matter contribution
in the generating functional of the Euclidean theory, one finds 
that the dominant contribution 
to the Euclidean path integral is given by
gravitational instantons 
(i.e. non-singular solutions of the 
Euclidean Einstein equations). 
In space-times with event horizons, this  
usually implies that metrics extremizing the Euclidean action are 
gravitational instantons only after removal of the conical singularity at the 
horizon\cite{GH}.
A period must therefore be fixed in the imaginary time, which 
becomes a sort of angular coordinate. 
It is well known that Euclidean quantum field 
theory with periodic imaginary time is equivalent to a finite-temperature 
quantum field theory in Lorentzian space-time, where the  
temperature is the inverse of the imaginary time period. In such a way
thermodynamics appears as a request of consistency of quantum 
field theory on space-times with Killing 
horizons, and in this sense we shall define such a thermodynamics 
as ``intrinsic".

In Refs. \cite{HHi,HHii} it has been shown that the 
Bekenstein--Hawking law, 
$S=A/4$, for
\BH\ entropy fails for extremal ones. These objects were already 
considered ``peculiar", since 
their metric does not show any conical structure near their event horizon, 
so no conical singularity removal is required.
The discovery of a zero entropy 
for extremal \BH s,  despite a non-zero 
area of the event horizon, made them even more important for the 
present 
investigation on the event horizon 
thermodynamics. 
In Ref. \cite{HHi} it has been observed 
that the source of such a different behavior 
between extremal and non-extremal \BH s is due to a change in the 
topological structure.  
In the latter case 
the presence of the event horizon is no longer 
associated with a 
non-trivial topology; 
the Euler characteristic 
indeed vanishes for extremal \BH s,  whereas it 
is different from zero for the 
non-extremal ones. All 
these considerations seem 
to  suggest 
that extremal black holes 
should be considered as a rather 
different object from the non-extremal ones. 

In this  work we shall 
prove that Euler characteristic and 
gravitational entropy can be related in the same way in almost all known 
gravitational instantons endowed with event horizons.
In particular,  
we shall show that the Euler characteristic and entropy
have the same dependence on the boundaries of the manifold and
we will relate them by a general formula. 
This formulation extends to a wide class 
of instantons, and in particular to the Kerr metric, the known results 
\cite{GHI,GK,BTZ,TC} about such a dependence. 

Finally, it is important to stress that in order to obtain this result one 
has to consider not only the manifold $M$, 
associated to the Euclidean section 
describing the instanton, but also the related manifold $V$, which 
is bounded  by 
the sets of fixed points of the Killing vector, associated 
to isometries in the imaginary time.
This should imply boundary contributions also for
cosmological, compact solutions.

\section{Euler characteristic and manifold structure}
\label{sec:due}

The Gauss--Bonnet theorem proves that it is possible to obtain
the Euler characteristic of a closed Riemannian manifold $M^n$ without
boundary from the volume integral of the four-dimensional curvature: 
\begin{equation}
S_{\GB}={{1}\over{32 \pi^{2}}} \int_{M} \varepsilon_{abcd}
R^{ab}\wedge R^{cd} ,\label{egb}
\end{equation}
where the curvature two-form ${R^a}_b$ is defined by the spin 
connection one-forms ${\omega^a}_{b}$ (for details see the Appendix) as 
\begin{equation}
{R^a}_{b}=d{\omega^{a}}_{b}+{\omega^a}_{c}\wedge
{\omega^c}_{b}.
\label{Rab}
\end{equation}
In a closed  Riemannian manifold $M^n$, Chern \cite{CHI,CHII} 
has defined  
the Gauss--Bonnet differential $n$-form $\Omega$ (with $n$ even) 
\begin{equation}
\Omega={{(-1)^{n/2}}\over{2^{n}\pi^{n/2}(\frac{n}{2})!}} \varepsilon_{a_{1}
\ldots a_{n/2}} R^{a_{1} a_{2}} \wedge \ldots \wedge
R^{a_{n+1} a_{n}}, 
\label{gaussint}
\end{equation}
and he has then shown that $\Omega$ can be defined in a manifold $M^{2n-1}$
formed by the unit tangent vectors of $M^n$. In such a way 
$\Omega$ can be expressed as the exterior
derivative of a
differential $(n - 1)$-form in $M^{2n-1}$
\begin{equation}
\Omega=-d\Pi.
\end{equation}
He has also proved that the original integral of 
$\Omega$ over
$M^{n}$ can be
performed over a submanifold $V^{n}$. This $n$-dimensional 
submanifold is obtained as the image in $M^{2n-1}$ of a continuous 
unit tangent vector field defined 
over $M^n$ with some isolated singular points. 
By applying Stokes' theorem one thus gets 
\begin{equation}
S^{\text {volume}}_{\text {GB}}=\int_{M^{n}}\Omega=\int_{V^{n}}\Omega=
\int_{\partial V^{n}}\Pi.
\end{equation}
Since the boundary of $V^n$ corresponds exactly to the singular points 
of the continuous unit tangent vector field defined over $M^n$, and  
bearing in mind that 
the sum of the indices of a vector field is equal to the Euler 
characteristic, one finds that 
the integral of $\Pi$ over the boundary of $V^n$ is equal to 
the Euler number $\chi$.  
For manifolds with a boundary, this formula can be
generalized \cite{EGH}:
\begin{eqnarray}
S_{\GB}&=&S^{\text {volume}}_{\text {GB}}+S^{\text {boundary}}_{\text {GB}}
\nonumber\\
&=&\int_{M^{n}}\Omega-
\int_{\partial M^{n}}\Pi=\int_{\partial V^{n}}\Pi-
\int_{\partial M^{n}}\Pi .
\label{Sgb}
\end{eqnarray}
Thus, 
the Euler characteristic of 
a manifold 
$M^{n}$ vanishes when its boundary 
coincides with that of a submanifold 
$V^{n}$ of $M^{2n-1}$.

The four-manifolds under
consideration can have a boundary formed by two disconnected 
hypersurfaces, say $\partial M^{n} =(r_{in},r_{out})$. 
As far as $V^n$ is concerned, the above quoted 
unit tangent vector field coincides (again for the cases considered here) 
with the time-like Killing vector field $\partial/\partial\tau$.
Hence the boundary will be the fixed-point sets 
of such a  vector field. The event horizon is always such a set; 
then the boundaries of $V^n$ will be at $r_{h}$ and possibly 
at one of the 
actual boundaries of $M^{n}$ which, for sake of simplicity, we shall assume 
at $r_{out}$. 

\section{Entropy for manifolds with a boundary}
\label{sec3}

Following the definition of gravitational entropy adopted 
in Ref. \cite{KOP}, we consider a thermodynamical system with conserved
charges $C_{i}$ and relative potentials $\mu_{i}$, 
and  we then work in a 
grand-canonical ensemble. 
The grand-partition function $Z$, the free energy $W$ and 
the entropy $S$ are:
\begin{equation}
Z ={\rm{Tr}}\; \exp{[-(\beta H -\mu_{i} C_{i})]} 
=\exp{[-W]},
\end{equation}
\begin{equation}
W =E-TS-\mu_{i} C_{i},
\end{equation}
\begin{equation}
S=\beta(E-\mu_{i}C_{i})+\ln Z,
\end{equation}
respectively. 
At the tree level of the semiclassical expansion: 
\begin{eqnarray}
Z& \sim &\exp{[-I_{E}]}\nonumber\\
I_{E}&=&{{1}\over{16 \pi}} 
\int_{M}[(-R+2\Lambda)+L_{matter})]+{{1}\over{8\pi}}
\int_{\partial{M}} \left[ K \right], 
\label{ie}
\end{eqnarray}
where $I_{E}$ is the on-shell Euclidean action and $[K]=K-K_{0}$ is the 
difference between 
the extrinsic curvature of the manifold and that of 
a reference background. 
We want to stress here that the
procedure we just showed has some subtleties that are 
well studied in the
literature. One of these is that the \sch\ solution is 
a maximum and not a minimum of the Euclidean effective action. This
is related to the fact that any black hole in vacuum is a 
highly unstable object (as can be seen by the negative value of its
specific heat $c=-8 \pi M^2$); but of course any self-gravitating
system shows such a behavior, due to the attractive nature of gravity.
The problem of performing a  
thermodynamical analysis 
of black holes, considering a grand-canonical ensemble, 
has been studied thoroughly by York \cite{Yo}, who suggested to
consider the black hole in a box. Such a choice automatically
stabilizes the \BH\ and enables one to perform further semiclassical
calculation. One can also consider the higher-order corrections to
the action. Unfortunately neither one-loop gravitons contributions 
\cite{GPJ}
nor matter ones \cite{LB} seem to be able to stabilize the black hole, 
since they are 
small in comparison with the tree-level term, at least in the regime of
negligible back reaction (that is far away from the 
quantum gravity regime).

To compute $Z$ and $I_{E}$ it is important to correctly 
take into account the
boundaries of the manifold $M^4$. 
We now evaluate separately the two terms occurring in the 
right-hand side of Eq. (\ref{ie}).
To obtain $\beta(E-\mu_{i}C_{i})$ one can consider the 
probability of transition between two 
hypersurfaces at $\tau$ equals constant 
(where $\tau=i t$), say $\tau_{1}$ and $\tau_{2}$.
In the presence of conserved charges one gets \cite{KOP}:
\begin{equation}
\langle \tau_{1} | \tau_{2} \rangle =\exp{[-(\tau_{2}-\tau_{1})(E-
\mu_{i} C_{i})]}
\approx \exp
{
\left[-{I_{E}}\right]_{\partial V}}.
\label{ener}
\end{equation}

The last equality in this equation is explained by the fact that
a hypersurface at $\tau={\text{const}}$ has a
boundary corresponding to the sets of fixed points for 
the Killing vector $\partial/\partial \tau$. 
Hence its boundary coincides with that of $V^{n}$.

Remarkably, for the manifolds under consideration,  
$V_{\text{bulk}}=M_{\text{bulk}}$; therefore 
the bulk part of the entropy always
cancels also for metrics that are not Ricci-flat. 
The entropy then depends on boundary values of the 
extrinsic curvature only. Thus, one obtains
\begin{eqnarray}
S&=&\beta(E-\mu_{i} C_{i}) +\ln{Z}\nonumber\\
&=&{{1}\over{8 \pi}} \left (\int_{\partial V}
[K] - \int_{\partial M} [K ] \right)
. 
\label{entr}
\end{eqnarray}
The analogy between Eq. (\ref{entr}) and Eq. 
(\ref{Sgb}) is self-evident.
For the boundaries 
of $V$ and $M$, the same considerations as at the end 
of Sec. II hold.

\section{Gravitational entropy and Euler characteristic for spherically 
symmetric metrics}

In this section we will find, for a given class of 
Euclidean spherically symmetric 
metrics, a general relation between gravitational entropy and Euler 
characteristic. We will then explicitly treat the most interesting cases.

\subsection{Euler characteristic}

In this section we compute the Euler characteristic for 
Euclidean spherically symmetric metrics of the form
\begin{equation}
d s^{2}=e^{2U(r)}d t^{2}+e^{-2U(r)}d r^{2}+R^{2}(r)d^{2}\Omega.
\label{metr}
\end{equation}
The associated spin connections read
\begin{eqnarray}
\omega^{01}&=&{{1}\over{2}} ( e^{2U})^{\prime} dt,\qquad
\omega^{21}=e^{U}R^{\prime}d\theta, \nonumber\\
\omega^{31}&=&e^{U}R^{\prime}\sin \theta d \phi, \qquad
\omega^{32}=\cos \theta d \phi ,
\label{spc}
\end{eqnarray}
and the Gauss--Bonnet action takes the form \cite{GK} 
\begin{eqnarray}
S^{\text {volume}}_{\text{GB}}&=&
{{1}\over{32 \pi^{2}}} \int_{M} \varepsilon_{abcd}
R^{ab} \wedge R^{cd}={{1}\over{4 \pi^{2}}} \int_{V}
d(\omega^{01} \wedge R^{23})
\nonumber\\
&=&{{1}\over{4 \pi^{2}}}
\int_{\partial V} \omega^{01} \wedge R^{23} . 
\label{gbvo}
\end{eqnarray}
The boundary term is \cite{GK,EGH}
\begin{eqnarray}
S^{\text {boundary}}_{\text{GB}}&=&
-{{1}\over{32 \pi^{2}}} \int_{\partial M}
\varepsilon_{abcd} (2\theta^{ab} \wedge R^{cd}-{{4}\over{3}} 
\theta^{ab} \wedge \theta^{a}_{e} \wedge \theta^{eb})
\nonumber\\
&=&-{{1}\over{4 \pi^{2}}} \int_{\partial M}
\omega^{01} \wedge R^{23}.
\label{gbbo}
\end{eqnarray}
Combining Eqs. (\ref{gbvo}) and (\ref{gbbo}) one eventually gets
\begin{eqnarray}
\label{GBSS}
S_{\GB}&=&S^{\text{volume}}_{\text{GB}}+S^{\text{boundary}}_{\text{GB}}\cr
&&\cr
&=&{{1}\over{4 \pi^{2}}} \left ( \int_{\partial V}-\int_{\partial M} \right )
\omega^{01} \wedge R^{23}.
\end{eqnarray}
where for the metrics (\ref{metr}) 
\begin{eqnarray}
R^{23}&=&d\omega^{23}+\omega^{21} \wedge \omega^{13}=(1-
e^{2U}(R^{\prime})^{2}) d\Omega \nonumber\\
\omega^{01} \wedge R^{23}&=&{{1}\over{2}} ( e^{2U} )^{\prime}
(1-e^{2U}(R^{\prime})^{2}) d\Omega\;  d t ,
\end{eqnarray}
and $d\Omega \equiv \sin\theta d\theta d\phi$ is the solid angle.
\pano
As already said, we perform our calculations on
Riemannian manifolds
with compactification of imaginary time, $0 \leq\tau
\leq\beta$, which is the 
generalization of the conical singularity removal condition for
the metrics under consideration.
It is easy to see that this corresponds to choosing
\footnote{Note that condition (\ref{b}) gives an infinite range of time (no
period) for extremal \BH\ metrics
(i.e. $\left.  (e^{2U})^{\prime}\right|_{r=r_{h}} =0$).
This leaves open the question to know 
if the period of imaginary time remains 
unfixed or if  
it has to be infinite, in correspondence to a zero temperature 
\cite{HHi}.}
\begin{equation}
\beta=4 \pi \left ( (e^{2U})^{\prime}_{r=r_{ h}} \right )^{-1}.
\label{b}
\end{equation}
By expressing Eq. (\ref{GBSS}) as a function of 
the actual boundaries, which are $\partial {V^4}=(r_{h},r_{out})$
and $\partial
{M^4}=(r_{ in},r_{out})$,
one gets
\begin{eqnarray}
S_{\GB}&=&2\left[1-(e^{U} R^{\prime})^{2} \right]_{r_{h}}\nonumber\\
&&\nonumber\\
&&-\left( (e^{2U})^{\prime}_{r=r_{h}} \right )^{-1}
\left[ (e^{2U})^{\prime} 
(1-(e^{U} R^{\prime})^{2} )\right]_{r_{in}}.
\label{Sgbg}
\end{eqnarray}
We can also rewrite Eq. (\ref{Sgbg}) in a more suitable form for our
next purposes:
\begin{equation}
\chi={{\beta}\over{2\pi}} \left[(2 U^{\prime} 
e^{2U})(1-e^{2U}{R^{\prime}}^{2})\right]^{r_{h}}_{r_{in}}, 
\label{chib}
\end{equation}
expressing the Euler characteristic as a function of the inverse
temperature $\beta$.

\subsection{Entropy}
\label{sec5}
For the metrics (\ref{metr}) one can obviously use the general formula 
(\ref{entr}). It is well known \cite{KOP} that one can write 
\begin{equation}
[K ]=\int_{\partial M} [\omega^{\mu} n_{\mu}],
\label{Ko}
\end{equation}
where for the metrics (\ref{metr}) under investigation, 
$\omega^{\nu}$ 
and $n_{\nu}$ are 
\begin{eqnarray}
\omega^{\mu}&=&\left(0,-2e^{2U}\left(\partial_{r} U+
2 \partial_{r} \ln R \right),
-{{2 \cot \theta}\over {r^{2}}}, 0 \right), \nonumber \\
n_{\mu}&=&\left( 0, {{1}\over{\sqrt{g^{11}}}},0,0 \right), 
\label{om}
\end{eqnarray}
and they lead to
\begin{equation}
\omega^{\mu}n_{\mu}=\omega^{1}n_{1}=-2
e^{U}\left (\partial_{r} U+2\partial_{r} \ln R \right).
\label{omc}
\end{equation}
By subtracting from Eq. (\ref{omc}) the flat metric
correspondent term
\begin{equation}\label{flat}
d s^{2}=d t^{2}+d r^{2}+r^{2} d\Omega^{2},
\end{equation}
one obtains
\begin{eqnarray}
{\omega^{\mu}_{0}}&=&\left(0,-{4\over r},-
{{2\cot \theta} \over{r^{2}}},0 \right),\nonumber \\
{n^{0}_{\mu}}&=&(0,1,0,0) ,
\label{os}
\end{eqnarray}
and
\begin{eqnarray}
[\omega^{\mu}n_{\mu}]&=&\omega^{\mu}n_{\mu}-
\omega_{0}^{\mu}n^{0}_{\mu}\nonumber \\
&=&-2e^{U}(\partial_{r}U+2\partial_{r}\ln R)+{{4}\over{r}} .
\label{of}
\end{eqnarray}
Performing  the integration of Eq. (\ref{entr}) for a 
spherically symmetric metric, and 
writing explicitly the dependence on boundaries, one gets
\begin{equation}
S=-\left. {{\beta R}\over{2}} \left[
(U^{\prime}R+2R^{\prime})e^{U}-{{2R}\over{r}} \right] e^{U}
\right|^{r_{in}}_{r_{h}}.
\label{eg}
\end{equation}

\subsection{Entropy and Topology}

We now prove that a relation between the gravitational entropy 
and the Euler characteristic can be found for the general case under 
consideration. One has
\begin{eqnarray}
A &=&4 \pi R^2 (r_h)\nonumber\\
\beta &=&4\pi((e^{2U})'_{r=r_h})^{-1}\nonumber\\
S &=& \left.{{\beta R}\over{2}} \left[(U'R+2R')e^{U}-
{{2R}\over{r}}\right]e^{U}\right|_{r=r_h}\nonumber\\
\chi &=&\left.{{\beta}\over{2 \pi}}(2U' e^{2U})(1-e^{2U}R'^{2})\right|_{r=r_h};
\label{gene}
\end{eqnarray}
hence one can relate $S$ and $\chi$ by their common dependence on $\beta$
\begin{equation}
S=\left.\frac{\pi \chi R}{ \left( 2U' e^{2U} \right) 
\left(1-e^{2U}R'^{2}\right)}
\left[(U'R+2R')e^{2U}-{{2R}\over{r}}e^{U}\right]\right|_{r=r_h}.
\label{gen1}
\end{equation}
By definition one has $\left. e^{2U}\right|_{r=r_h}=0$, and  
Eq. (\ref{gen1}) then yields
\begin{equation}
S = \pi \chi R(r_h) \left. \left [ (e^{2U})' \right]^{-1} \right|_{r=r_h}
={{\pi \chi R^{2}(r_h)}\over{2}}=
{{\chi A}\over{8}} .
\label{SC}
\end{equation}
Some remarks on Eq. (\ref{SC}) are in order. 
We evaluated Eq. (\ref{entr}) in a grand-canonical ensemble 
so  this formula {\it a priori} is valid only for instantons endowed 
with non-zero temperature. 
Nevertheless, as we said, for extremal \BH s there is no conical 
singularity, and therefore no $\beta$ fixing. 
The fact that Eq. (\ref{SC}) gives the expected result 
also for extremal solutions (as seen in these cases one gets $\chi=0$, 
which 
in Eq. (\ref{SC}) straightforwardly gives $S=0$)~\footnote{We 
are referring here to the semiclassical
results\cite{HHi,HHii} quoted above. 
For a discussion about the discrepancies w.r.t. string theory
calculations, see the Conclusions.} enables us to 
conjecture that Eq. (\ref{SC}) is the general formula, 
which can be applied to all the known cases of instantons with horizons. 
The eventual lack of intrinsic thermodynamics is simply
deducible from Eq. (\ref{SC}) by considerations about the topology of the 
manifold. 

We will prove this assumption by studying most of the known  
solutions with intrinsic thermodynamics.
We will start with metrics of the (\ref{metr}) form.
Moreover, we will show that 
Eq. (\ref{SC}) also holds for an instanton with an Euler 
characteristic different from 2 (i.e. the Nariai one where
the Euler 
characteristic equals 4) and for an axisymmetric one (i.e. 
the Kerr metric) 
which cannot be cast in the form of Eq. (\ref{metr}).

We will consider both black hole and cosmological solutions.
The formers are asymptotically flat 
solutions, hence they always have a boundary at
infinity, $r_{out}=\infty$. The inner boundary of 
$M$ is usually missing since the horizon, after removal of
the conical singularity, becomes a regular point
of the manifold.  
By contrast, a drastic change in the 
boundary structure occurs for extremal \BH s.
In such a case we cannot fix
imaginary time
value since metrics present no conical singularity. 
The horizon is at infinite distance from the external
observer; hence 
it is an inner boundary of $M^4$ (i.e. the
coordinate of this inner boundary is $r_{in}$).

As far as the cosmological solutions are concerned, they are compact, 
and therefore 
$\partial M=0$. Instead, the boundary of $V^{n}$ is only at the 
horizon that now is also the maximal radius for the space; hence 
the formulas 
for entropy and Euler characteristic are still applicable, setting 
$r_{out}=
0$ and reversing the sign in front of the equations.

\section{Spherically symmetric metrics}

\subsection{\sch\ instanton}

We first consider the \sch\ \BH , 
where 
\begin{eqnarray}
e^{2U}&=&(1-2M/r),
\nonumber \\
U&=&{1\over 2 }\ln (1-2M/r),\nonumber \\
R&=&r . 
\end{eqnarray}
Using Eq. (\ref{b}) and bearing in mind that 
$A=\beta r_{h}=4 \pi r^{2}_{h}$,
one can write the relation
between
$\beta$ and $A$ as 
\begin{equation}
\beta={{A}\over{r_{h}}}.
\label{inv}
\end{equation}
Moreover, from Eq. (\ref{eg}) one gets
\begin{equation}
S={{A}\over{4}},
\label{Ss}
\end{equation}
and from Eq. (\ref{chib}) one also finds   
\begin{equation}
\chi=\beta r_{h} {{1}\over {2\pi r^{2}_{h}}} ={{A}\over{2 \pi
r^{2}_{h}}}.
\label{Cs}
\end{equation}
Now, combining Eqs. (\ref{Ss}) and (\ref{Cs}), one 
obtains
\begin{equation}
S={{\pi}\over{2} } \chi r^{2}_{h}={{\chi}\over{32
\pi}}\beta^{2}={{\chi A}\over{8}}.
\end{equation}

\subsection{Dilaton U(1) black holes}

In the case of the dilaton $U(1)$  \BH\ solutions parametrized by $0 \leq a 
\leq 1$ (where $a=0$ corresponds to the Reissner--Nordstr\"om \BH) one has
\begin{eqnarray}
e^{2U}&=&\left (1-{{r_{+}}\over{r}} \right )\left (1-{{r_{-}}
\over{r}} \right )^{{{1-a^{2}}\over{1+a^{2}}}}
\nonumber \\
U&=&{1\over 2 }\ln \left [\left (1-{{r_{+}}\over{r}} \right )\left
(1-{{r_{-}}\over{r}} \right )^{{{1-a^{2}}\over{1+a^{2}}}}\right]
\nonumber \\
R&=&r \left (1-{{r_{-}}\over{r}} \right )^{{{a^{2}}\over{1+a^{2}}}}
\nonumber \\
M&=&{{r_{+}}\over{2}}+{{1-a^{2}}\over{1+a^{2}}}{{r_{-}}\over{2}}
\nonumber \\
Q^{2}&=&{{r_{+}r_{-}}\over{1+a^{2}}} 
\nonumber \\
r_{h}&=&r_{+}.
\end{eqnarray}
For such \BH s one finds $A=\beta R_{r_{h}}=4 \pi
R^{2}_{r_{h}}$, where 
$R$ determines the characteristic scale of distance. As before
(cf. Eq. (\ref{inv})):
\begin{equation}
\beta ={{A}\over {r_{h} \left ( 1- {{r_{-}}\over{r_{h}}} \right ) }}.
\end{equation}
From Eq. (\ref{eg}) one obtains 
\begin{equation}
S=4 \pi R^{2}_{r_{h}}={{A}\over {4}},
\label{Sd}
\end{equation}
and from Eq. (\ref{chib}) one finds
\begin{equation}
\chi=\beta r_{h} {{1}\over {2\pi R^{2}_{h}}} ={{A}\over{2 \pi
R^{2}_{h}}}.
\label{Cd}
\end{equation}
Again, combining Eqs. (\ref{Sd}) and (\ref{Cd}), one gets  
\begin{equation}
S={{\pi}\over{2} } \chi R^{2}_{h}={{\chi A}\over{8}}.
\end{equation}

\subsection{de Sitter instanton}
In the de Sitter cosmological case, we can prove 
how the relation Eq. (\ref{SC}) is 
due to the boundary structure (horizons and ``real" boundaries) of
the manifold and not to the presence of a black hole.
There now is only a cosmological horizon and no proper 
boundary for $M$, and the topology of the de Sitter instanton 
is a four-sphere. One has 
\begin{eqnarray}
e^{2U}&=&\left(1- {{\Lambda}\over{3}} r^{2}\right)
\nonumber \\
U&=&{1\over 2 }\ln \left(1- {{\Lambda}\over{3}}  r^{2}\right)
\nonumber \\
R&=&r
\nonumber \\
r_{\Lambda}&=&\sqrt{{3}\over{\Lambda}}
\nonumber \\
A &=& \frac{12\pi}{\Lambda}
\nonumber \\
\beta &=& 2\pi \sqrt{\frac{3}{\Lambda}}
\end{eqnarray}
where $r_{\Lambda}$ and $A$ are respectively the 
radius and area of the  cosmological horizon. 
For such compact manifold no Minkowskian subtraction is needed; 
hence, by using Eq. (\ref{omc})  
one straightforwardly gets 
\begin{equation}
S={{1}\over{8 \pi}} \int_{\partial V} K .
\label{cosmentr}
\end{equation} 
From Eq. (\ref{of}) it is easy to find 
\begin{eqnarray}
\omega^{\mu}n_{\mu}&=&
-2e^{U}\left(\partial_{r}U+2 \partial_{r} \ln R  \right )
\nonumber \\
&=&2\left({{r \Lambda}\over {3}}
{{1}\over{\left[1-\left({{r^{2}\Lambda}\over{3}}\right ) \right]^{1/2}}}
-{{2}\over{r}}
\left[1-\left({{r^{2}\Lambda}\over{3}}\right ) \right]^{1/2} 
\right)
\end{eqnarray}
Hence, bearing in mind Eq. (\ref{Ko}), one obtains
\begin{equation}
S={{1}\over{16 \pi}} \int_{r_{\Lambda}} 
\omega^{\mu}n_{\mu}
e^{U}\; r^2 \sin{\theta}\; d\theta\; d\tau \; d\phi = 
\frac{\beta^{2}}{4 \pi}.
\label{SdeS}
\end{equation}
By using Eq. (\ref{chib}) with $r_{in}=0$ 
and $r_{h}=r_{\Lambda}$, the Euler 
characteristic is 
\begin{equation}
\chi=\frac{\beta^{2} \Lambda}{6 \pi^{2}}.
\label{CdS}
\end{equation}
Combining then Eqs. (\ref{SdeS}) and (\ref{CdS}),  
it is easily 
checked that Eq. (\ref{SC}) also holds in the de Sitter case.   

\subsection{Nariai instanton}

The Nariai instanton is the only non-singular solution of Euclidean vacuum
Einstein equation for a given mass $M$ and cosmological constant $\Lambda$.
It can be regarded as the limiting case of the \sch --de Sitter solution 
when one equals the surface gravity of the \BH\ to that of the 
cosmological horizon in order to remove all conical singularities. 
This could seem meaningless since,   
in \sch --de Sitter 
coordinates, the Euclidean section shrinks to zero (the \BH\ and 
cosmological horizons coincide). However, on making an 
appropriate change of coordinates \cite{GP,HB}, the volume of the 
Euclidean section no longer vanishes, and the 
space-time can be properly studied. In this coordinate system, one 
still deals with a spherically symmetric metric, and the 
vierbein forms are
\begin{eqnarray}
e^{0}&=& {1 \over \sqrt \Lambda}\; \sin\xi  d\psi, \qquad
e^{1}={1 \over \sqrt \Lambda}\;d\xi,\\
e^{2}&=&{1 \over \sqrt \Lambda}\;d\theta, \qquad
e^{3}={1 \over \sqrt \Lambda} \;\sin\theta d\phi.
\end{eqnarray}
One then obtains
\begin{eqnarray}
de^{0}&=&-{\sqrt\Lambda}\cot\xi \; e^{0}\wedge e^{1} , \qquad
de^{1}=de^{2}=0,\cr
de^{3}&=&{\sqrt\Lambda}\cot\theta\; e^{2}\wedge e^{3},
\end{eqnarray}
and 
\begin{eqnarray}
R^{01}&=&d\omega^{01}=\Lambda\;  e^{0}\wedge e^{1},\\
R^{23}&=&d\omega^{23}=\Lambda\; e^{2}\wedge e^{3}.
\end{eqnarray}
Moreover one has 
\begin{eqnarray}
R &=& \Lambda^{-1/2} ,
\nonumber\\
A &=& {4\pi \over \Lambda},
\nonumber\\
\beta &=& \frac{2 \pi}{\sqrt {\Lambda}}.
\end{eqnarray}
The ranges of integration are 
$0\leq\psi\leq\beta\sqrt{\Lambda}$, 
$0\leq\xi\leq\pi$, 
$0\leq\theta\leq2\pi$,
$0\leq\phi\leq2\pi$.
The extremes of $\xi$  correspond to the cosmological 
horizon and to the \BH\ horizon \cite{HB}.
It is worth noting that the period of the imaginary time, $\psi$, 
is 
$\beta\sqrt{\Lambda}$, instead of the usual $\beta$. 
This is due to the normalization of the time-like 
Killing vector one is forced 
to choose in this space-time\footnote{For a wider discussion 
of this point, see 
the Appendix of Ref. \cite{HB}.}. 
The form of the Nariai metric does not enable us to apply Eq. (\ref{chib}) 
and  we then compute the Euler characteristic from the very beginning. 
We obtain
\begin{eqnarray}
\label{chinar}
S_{\rm GB}&=&{\Lambda^{2}\over 4\pi^2}
\int_{V} e^{0}\wedge e^{1}\wedge e^{2}\wedge e^{3}\\
\nonumber
&=&{1\over 4\pi^2} \int_{0}^{\pi} \sin\xi d\xi 
\int_{0}^{\beta\sqrt{\Lambda}}d\psi \int_{0}^{\pi} \sin\theta d\theta 
\int_{0}^{2\pi}d\phi\\ 
&=&\frac{2\beta\sqrt{\Lambda}}{\pi} .
\nonumber
\end{eqnarray}
By substituting $\beta$, one can check that Eq. (\ref{chinar}) gives 
the correct result. 
In fact, the Nariai instanton has topology $S^{2} \times S^{2}$; 
hence its Euler number, bearing in mind the product formula, is 
$\chi=2\times2=4$.

The entropy can be easily calculated from Eq. (\ref{cosmentr}).
In this case the extrinsic curvature is given by
\begin{equation}
K=-\sqrt{\Lambda}\;\frac{\cos \xi}{\sin\xi},
\end{equation} 
and one obtains
\begin{eqnarray}
S&=&-\frac{1}{8\pi}\int_{0}^{2\pi}d\phi\int_{0}^{\pi}\sin\theta d\theta 
\int_{0}^{\beta\sqrt{\Lambda}} \left[ 
\frac{\sqrt{\Lambda} \cos \xi}{\Lambda^{3/2}}\right]_{0}^{\pi} 
d\psi
\nonumber\\
&=&\frac{\beta}{\sqrt{\Lambda}}. 
\label{snar}
\end{eqnarray}
It is now easy to check that the combination of Eqs. (\ref{chinar}) and 
(\ref{snar}) gives Eq. (\ref{SC}).

Remarkably, this implies that Eq. (\ref{SC}) cannot be 
cast in the form 
\begin{equation}
S=\left(\frac{\chi}{2}\right)^{\alpha}\frac{A}{4},
\label{pro}
\end{equation}
where $\alpha$ could in principle be any positive constant.
Since Eq. (\ref{SC}) holds also 
for the Nariai instanton, $\alpha$ must be fixed to 1.

\section{Kerr metric}
\label{sec:Kerr}
The Kerr solution describes both the stationary 
axisymmetric asymptotically flat gravitational field 
outside a massive rotating body and a rotating 
\BH\ with mass $M$ and angular momentum $J$. The Kerr \BH\ 
can also be viewed as the final state of a collapsing star, 
uniquely determined by its mass and rate of rotation. 
Moreover, its thermodynamical behavior is 
very different from 
\sch\ or Reissner--N\"ordstrom \BH s, because of its 
much more complicated causal 
structure\footnote{For instance, 
Wald pointed out that in a Kerr \BH\ it is not 
possible to mimic the Unruh--Rindler case to explain its thermal 
behavior\cite{Wald}.}. Hence its study is of great interest in understanding  
physical properties 
of astrophysical objects, as well as  
in checking any conjecture about thermodynamical 
properties of \BH s.

In terms of Boyer--Lindquist coordinates, the Euclidean Kerr metric 
reads \cite{Oneill}
\begin{eqnarray}
\label{kermet}
ds^{2} &=& {\Delta \over \rho^{2}}
\left[dt-a{\sin^{2}\!\theta}d\varphi\right]^{2}
+ {\rho^{2} \over \Delta}dr^{2}\cr
&&+\rho^{2}d\theta^{2} 
+{\sin^{2}\theta \over \rho^{2}}\Bigr[\left(
r^{2}+a^{2}\right)d\varphi -adt\Bigr]^{2},
\end{eqnarray}
where
\begin{eqnarray}
\rho &= &r^{2} + a^{2}\cos^{2}\theta, \nonumber\\ 
\Delta &=& r^{2}-2Mr +a^{2}.
\end{eqnarray}
Here $a$ is the angular momentum 
for unit mass as measured from the infinity; it 
vanishes in the \sch\ limit, and $\Delta$ is the Kerr 
horizon function. 
The roots of the horizon function 
$\Delta$ correspond to two null-like surfaces at 
\begin{equation}
r_{\pm}=M \pm \sqrt{M^2-a^2},
\end{equation}
where $r_{+}$ is the Kerr \BH\ event horizon
and $r_{-}$ is 
the Cauchy horizon around the ring singularity at $\rho=0$.  
The area and the \BH\ angular velocity are 
respectively
\begin{eqnarray}
A &=& 4\pi(r_{h}^{2}+a^{2}),\\
\Omega &=& \frac{a}{(r^{2}_{h}+a^{2})},
\label{AO}
\end{eqnarray}
Such a metric corresponds to the following vierbein forms: 
\begin{eqnarray}
\label{vierb}
e^{0}&=&{{\sqrt\Delta} \over \rho}\left(dt-a\sin^{2}\!\theta
d\varphi\right), \qquad
e^{1}= {\rho \over{\sqrt\Delta}}dr, \\
e^{2}&=&\rho d\theta, \qquad\quad e^{3}=
{\sin\theta\over \rho}
\Bigr[\left(r^{2}+a^{2}\right)d\varphi -adt\Bigr],
\end{eqnarray}
where $\rho$ is the positive square root of $\rho^2$.

From Eq. (\ref{vierb}), one can obtain the spin connection one-forms 
as \cite{Oneill}
\begin{eqnarray}\label{kerrform}
{\omega^{0}}_{1}&=&
-{ar\sin\theta\over {\rho^{3}}}
e^{3} + Fe^{0}, \nonumber \\
{\omega^{0}}_{2}&=&
-{{a\cos\theta{\sqrt\Delta}} \over {\rho^{3}}}e^{3}
-a^{2}
{\sin\theta\cos\theta \over \rho^{3}}e^{0}, \nonumber \\
{{\omega^{0}}_{3}}&=&
 -{ar\sin\theta\over \rho^{3}}
e^{1}+{a\cos\theta{\sqrt\Delta} \over \rho^{3}}e^{2} ,
\nonumber\\
{\omega^{1}}_{2}&=&
-a^{2}{\sin\theta\cos\theta
\over \rho^{3}}e^{1}-
r{{\sqrt\Delta}\over\rho^{3}}e^{2},\nonumber\\
{\omega^{3}}_{1}&=&-{\omega^{1}}_{3}=
r{{\sqrt\Delta}\over\rho^{3}}e^{3}
 +{ar\sin\theta\over \rho^{3}}
e^{0},\nonumber\\
{\omega^{2}}_{3}&=&
-{\cos\theta \over \sin\theta}{{\left(r^{2}
+a^{2}\right)}\over \rho^{3}}e^{3}
-{a\over \rho^{3}}\cos\theta{\sqrt\Delta}e^{0},
\end{eqnarray} 
where \cite{Oneill}
\begin{equation}
F\equiv{\partial \over \partial r} {{\sqrt \Delta}
\over \rho}={{\left(r-M\right)\rho^{2}-r\Delta}
\over \rho^{3}{\sqrt\Delta}}.
\end{equation}
By virtue of Eqs. (\ref{kerrform}) and (\ref{Rab}), and
the nilpotency of the exterior derivative operator $d$,
the Gauss--Bonnet action in Eq. (\ref{Sgb}) takes the form
\begin{eqnarray}
\label{kerraction}
S_{\text{GB}}=
-{{1}\over{4 \pi^{2}}}\int &&\left(\omega^{01} \wedge R^{23}
+\omega^{02}\wedge \omega^{21}\wedge\omega^{23}\right.\cr 
&&\left.+\omega^{03}\wedge \omega^{31}\wedge\omega^{23}
+\omega^{02}\wedge d\omega^{31} \right)_{r_h}, 
\end{eqnarray}
where $d\omega^{31}$ can be expressed in terms of 
a suitable combination (wedge product) of the type $e^{i}\wedge
e^{j}$, and $r_h$ is the radius of the Kerr horizon (i.e. 
the positive 
roots of $\Delta=0$). 
For further details see the Appendix and Ref. \cite{Oneill}.
By defining the following quantities
\cite{Oneill}:
\begin{eqnarray}
I&\equiv&{Mr \over {\rho^6}}\left(r^{2} -3a^{2}\cos^{2}\theta\right),\cr
&&\\ 
K&\equiv&{Ma\cos\theta \over \rho^{6}}\left(3r^{2} -a^{2}
\cos^{2}\theta\right),
\nonumber
\end{eqnarray}
one obtains, for the quantity in round brackets, in
Eq. (\ref{kerraction}):
\begin{eqnarray}
&& e^{0}\wedge e^{2}\wedge e^{3} \;
\left(2FI  {3ra^{4}(\sin\theta 
\cos\theta)^{2}\sqrt{\Delta}\over 
\rho^{9}} \right)
\nonumber\\
&-& e^{0}\wedge e^{1}\wedge e^{3} \;
\left(
{8Mr^{3}a^{2}\sin\theta \cos\theta \over \rho^{9}}
\right),\nonumber
\end{eqnarray}
in terms of one-forms appearing in Eq. (\ref{vierb}).
Such a quantity has to be evaluated at $r=r_h$. 

At this stage some remarks are in order.
In the Euclidean path-integral approach the Kerr solution is an instanton 
(i.e. a non-singular solution of the Euclidean action)
only after the 
identification of the points $(\tau,r,\theta,\varphi)$ and 
$(\tau+2\pi\kappa_{1}^{-1},r,\theta,\varphi+2\pi \kappa_{1}^{-1}\kappa_{2})$ 
\cite{GH}, where $\kappa_{1}=\kappa$ is the surface gravity of the \BH\ 
and $\kappa_{2}=\pm \Omega$. 
With this identification, the Euclidean section has topology $R^2\times 
S^2$ and  
$\chi=2$. The condition of a periodic isometry group 
implies  $\kappa_{2}/\kappa_{1}=q$ \cite{GHI}, where $q \in Q$ is a rational 
number. By using this relation, it is easy to see that the periods are: 
\begin{eqnarray}
\beta_{\tau} &=& 2\pi \kappa_{1} = 4\pi 
\frac{Mr_{ h}}{\sqrt{(M^2-a^2)}},\nonumber\\
\beta_{\varphi} &= & 2\pi \frac{\kappa_{1} }{k_{2}}= 2\pi q,
\label{betak}
\end{eqnarray}
If one would set $q \neq 1$, Eq. (\ref{entr}) for the \BH\ entropy
would acquire a factor $q$, but this spurious factor would be absorbed
in the change of the period of $\varphi$ that implies 
a redefinition of the \BH\ area (\ref{AO}), which 
would become $A=4\pi q(r_{ h}^{2}+a^2)$. 
Therefore one still expects $S=A/4$, and the fixing of $q=1$ will not bring
about a loss of generality. 
Moreover in this way the area will be the ``physical'' one, as
written in Eq. (\ref{AO}). Hence the Euler number is
\begin{eqnarray}
\chi &=&
\frac{M{r_{ h}}({r_{ h}}-M)}{4\pi^{2}}
{\int_{0}^{\beta}}d\tau
{\int_{0}^{2\pi}}
{d\varphi} {\int_{0}^{\pi}}
{
{\left(r_{ h}^{2}-3a^{4}\cos^{4}\!\theta\right)} 
\over {\left(r_{h}^{2}+a^{2}\cos^{2}\!\theta\right)^{3}}
}
\sin\theta d\theta \nonumber\\
&=&{2 \over \pi} \beta (r_{h}-M) {{Mr_{h}} 
\over  {\left(r_{h}^{2}+a^{2}\right)^{2}}}.
\label{eq7}
\end{eqnarray}
Bearing in mind Eq. (\ref{betak}) 
and that $(r_{h}^{2}+a^{2})= 2Mr_{h}$, one eventually gets  
\begin{equation}
\chi =8 {M^{2}r_{h}^{2}\over 
\left(r_{h}^{2}+a^{2}\right)^{2}}= 2.
\end{equation}

As far as the entropy is concerned, we here follow the procedure outlined 
in Sec. \ref{sec3} and \ref{sec5}. From Eq. (\ref{entr}), writing 
$\omega^{\mu}$ as 
\begin{equation}
\omega^{\mu}= -{2 \over \sqrt{g}}
\left({\partial {\sqrt g}\over \partial x^{\nu}}\right)
g^{\mu\nu} -{\partial g_{\nu\mu}\over \partial x^{\nu}},
\end{equation} 
and bearing in mind that the Kerr determinant is
\begin{equation}
{\sqrt g} = \rho^{2} \sin\theta ,
\end{equation}
one finds
\begin{eqnarray}
\omega^{\mu}&=& \left(0,-2{r\Delta\over \rho^{4}},
-{2(r-M)\over \rho^{2}}, -2{\cot\theta\over \rho^{2}},0\right),
\nonumber\\
n_{\mu} &=&\left(0,{\rho \over {\sqrt\Delta}},0,0\right).
\end{eqnarray}
By subtracting the flat Minkowskian term of $\omega^\mu$ (see Eq. (5.7))
one easily obtains  
\begin{equation}
\left[\omega^{\mu}n_{\mu}\right]
=-{2 \over \rho {\sqrt\Delta}} \left(
{r\Delta \over \rho^{2}} + r-M\right) 
+ {4\over r}.
\end{equation}
One  can then evaluate the Kerr 
\BH\  entropy: 
\begin{eqnarray} 
S&=&-{1\over 16\pi} \int_{0}^{\beta} d\tau
\int_{0}^{\pi}d\varphi\int_{0}^{2\pi}d\theta
\rho{\sqrt \Delta}\sin\theta \:\:\cdot\cr
&&\left[-{2 \over \rho {\sqrt\Delta}} \left(
{r\Delta \over \rho^{2}} + r-M\right)
+ {4\over r}\right]_{r_h}= {\beta \over 2} (r_{h} -M),
\label{equ19}
\end{eqnarray}
Thus, combining Eqs. (\ref{eq7}) and (\ref{equ19}), one has
\begin{equation}
S = {\pi \over 4}{\left(r_{h}^{2}+a^{2}\right)^{2}
\over Mr_{h}}\chi = {1\over 2}\pi \left(r_{h}^{2}+a^{2}\right) \chi=
{A \over 8}\chi.
\end{equation}  
\section{CONCLUSIONS}
The main result of this work is a new formulation of the Bekenstein--Hawking 
entropy. This has been achieved by making explicit 
how gravitational entropy depends on topology.   
This result has been proved to be valid for a wide class of 
gravitational instantons endowed with intrinsic thermodynamics.
Therefore, it can be considered a confirmation and generalization 
of previous results \cite{BTZ,TC} (obtained for \BH s and in a different 
formalism) as well as a compact, general formulation of 
the Bekenstein--Hawking relation.

Although our results seem to imply a central role for space-time
topology in the explanation of 
intrinsic thermodynamics of gravitational
instantons, we are not claiming that it is not necessary to understand
the microscopic degrees of freedom of black holes in order to understand
their entropy. 
The fact that the horizon's area is still present in Eq. 
(\ref{SC}) implies
a dependence for the gravitational entropy on two different objects: a
discrete topological parameter (i.e. the Euler characteristic, which can be
$0,2$ or $4$ for the known solutions with event horizons) and the area,
which can vary with continuity. 
Since the topological term comes out in the general form (\ref{SC}),
for all the space-times endowed with intrinsic thermodynamics, this seems
to imply that the topological non-triviality of space-time is a necessary
(although probably not sufficient) condition for the coming up of the,
otherwise hidden, microscopic nature of gravity. The origin of these
hidden degrees of freedom is still a matter of debate. 
The authors suggest that the relation here 
clearly shown between the entropy and the boundary structure of the
four-manifold seems to add evidence in favor of an interpretation based
on the dynamical degrees of freedom associated to vacuum in topological
non-trivial four-manifolds\cite{mes}. 
This enables us to argue that intrinsic thermodynamics of some
gravitational instantons could be due to a sort of ``gravitational
Casimir effect'' \cite{BL} on such four-manifolds.

Finally we stress that the interesting results\cite{Sen,SV}, about
the non-zero entropy of some extremal string theory \BH\ solutions, are not
necessarily in contrast with ours.
In fact recent calculations\cite{GM2,GM} have proved that 
the discrepancies between semiclassical and string theory results can be
eliminated if one performs in the former approach a sum over topologies
and imposes the extremality condition after quantization.
It is easy to see\cite{GM} that with this procedure the non-trivial
topologies will dominate, in this case reducing the case to a non-extremal
one, with non-zero entropy. Hence it seems that string model results
implicitly involve a quantization procedure where the classical
extremal topology is ignored by means of the quantization procedure.
Of course it is still an open question which of the two procedures could better
fit reality.

In the opinion of the authors all these problems are deeply intertwined and 
hence they deserve further investigation.

\acknowledgements
S. Liberati wishes to thank F.~Belgiorno for 
illuminating remarks, M.~Maggiore and 
M.~Martellini for constructive advice.
G.~Pollifrone thanks D.~Bellisai for correspondence about Kerr black holes. 
Both authors are also grateful to P.~Blaga, G.~Esposito, 
R.~Garattini, G.~Immirzi, B.~Jensen  and 
K.~Yoshida for extensive discussions. The work of G.~Pollifrone 
was supported in part by the Angelo Della Riccia Foundation. 
\appendix
\section*{}

The Euler number of a four-manifold can be defined as an alternating sum
of Betti numbers:  
\begin{equation} \chi \equiv \sum_{n=0}^{4}
(-1)^{n}B_{n}. \end{equation}
The $n$th Betti number $B_n$ is the number of
independent closed $n$-surfaces that are not boundaries of some
$(p+1)$-surface.  For a compact manifold without boundary, $B_n$ is also
equal to the number of linear independent harmonic $n$-forms, and $B_{n}=
B_{4-n}$, (i.e. $B_{0}=B_{4}=1$ and $B_{1}=B_{3}$). If the four-manifold
is simply connected, $B_{1}=0$, whereas if there is a boundary, $B_{0}=1$
and $B_{4}=0$.

In the Cartan approach to geometry one deals with 
differential forms (see Sec.\ \ref{sec:due}). 
Defining a local coordinate basis of one-forms $dx^{\mu}$ 
and a local orthonormal basis of one-forms $e^{a}$ over a four-manifold 
$M$, 
the metric can be expressed as:
\begin{equation}
g = g_{\mu\nu}dx^{\mu}\otimes dx^{\nu}= \eta_{ab}e^{a}
\otimes e^{b} ,
\end{equation}
where $\eta_{ab}$ is the flat Euclidean metric tensor
with signature +4, and the information 
about the curvature of the Riemannian four-space is encoded 
in 
\begin{equation}
e^{b}= e^{b}_{\mu}dx^{\mu}. 
\end{equation}
Here $e^{a}_{\mu}(x)$ 
are the vierbein (or tetrad) one-forms, 
and they can be viewed as a sort of square root
of the metric. Note that the Greek letters $\mu,\nu,\ldots$ 
denote abstract indices, 
and latin letters $a,b,\ldots$ internal 
indices. 

We can now introduce the spin connection one-forms ${\omega^{a}}_{b}$ 
and define the first Cartan structure equation:
\begin{equation}
T^{a} \equiv de^{a}+{\omega^{a}}_{b}\wedge e^{b}
={1 \over 2} {T^a}_{bc} \; e^{b}\wedge e^{c},
\end{equation}
where $T^{a}$ is the torsion two-form of the manifold. 
The second Cartan 
structure equation defines the curvature two-form of the manifold
(see Eq. (\ref{egb})):
\begin{equation}
{R^a}_{b} \equiv d{\omega^a}_{b} + 
{\omega^a}_{c} \wedge{\omega^c}_{b} =
{1 \over 2} {R^a}_{bcd} \; e^{c}\wedge e^{d}.
\end{equation}
In the tensor formalism the covariant derivative $\nabla_{\alpha}$ is defined 
by using the Levi--Civita connection (or Christoffel symbols) 
$\Gamma^{\alpha}_{\mu\nu}$. 
By virtue of the 
metricity conditions (i.e. $\nabla_{\alpha}g_{\mu\nu}=0$)
and of the absence  of torsion (i.e. ${T^\mu}_{[\alpha\beta]}=0$),
the Levi--Civita connection is then uniquely determined in terms of the 
metric. 
In the Cartan approach, the spin-connection one-forms 
replace the Christoffel symbols. The Levi--Civita spin-connection
one-forms are then obtained  
by imposing the metricity and the torsionless conditions, which yield
\begin{equation}
\omega_{ab} =- \omega_{ba}
\end{equation}
and 
\begin{equation}
de^{a}+{\omega^{a}}_{b}\wedge e^{b}=0,
\end{equation}
respectively.\pano 
In Sec.\ \ref{sec:Kerr} the term $\varepsilon_{abcd}R^{ab}\wedge R^{cd}$
occurring in the Gauss--Bonnet action (\ref{kerraction}) reads 
\begin{eqnarray}
\varepsilon_{abcd}R^{ab}\wedge R^{cd}&=&
6\Bigr[d\left(\omega^{01} \wedge R^{23}+ \omega^{02}
\wedge \omega^{21}\wedge\omega^{23} \right.\cr 
&&\left.+\omega^{03}\wedge \omega^{31}\wedge\omega^{23}
+\omega^{02}\wedge d\omega^{31}
\right)\nonumber\\
&+& d\omega^{03}\wedge d\omega^{12}\Bigr].
\end{eqnarray}
The last term  $d\omega^{03}\wedge d\omega^{12}=
d(\omega^{03}\wedge d\omega^{12})$, bearing in mind 
that $e^{i}\wedge e^{i}=0$, vanishes by virtue 
of the structure of $\omega^{03}$ and $\omega^{12}$ (see 
Eq. (\ref{vierb})).
Furthermore, the term $d\omega^{31}$ takes the form:
\begin{eqnarray}
d\omega^{31}&=&-{a\sin\theta \over \rho^{3}}\left(rF + 
{\sqrt \Delta \over \rho^{3}}
a^{2}\cos^{2}\!\theta\right)e^{0}\wedge e^{1}
\nonumber\\
&&- {ar\cos\theta \over \rho^{6}}\Bigr(
\rho^{2} + 2 a^{2}\sin^{2}\!\theta\Bigr)e^{0}\wedge e^{2}
\nonumber\\
&&+ {1\over \rho^{6}}\Bigr[\Delta\left(\rho^{2}-2r^{2}\right)
+r\rho^{2}\left(r-M\right)\Bigr]e^{1}\wedge e^{3}
\nonumber\\
&&+ {r{\sqrt\Delta}\cos\theta \over \rho^{6}\sin\theta}
\Bigr(\rho^{2}- a^{2}\sin^{2}\!\theta\Bigr)e^{2}\wedge e^{3}.
\end{eqnarray}

\end{document}